\documentclass[aps,pra,preprint]{revtex4}
\usepackage{bm}
\usepackage{dcolumn}
\usepackage{amsmath}
\usepackage{times}
\newcolumntype{.}{D{x}{}{-1}}

\newcolumntype{w}[1]{D{.}{.}{#1}}

\begin{document}
\preprint{Version 1.2}

\title{Relativistic, QED, and nuclear mass effects in the magnetic
  shielding of $^3$He}

\author{Adam Rudzi\'nski}
\author{Mariusz Puchalski}
\email[]{mpuchals@fuw.edu.pl} 
\author{Krzysztof Pachucki}
\email[]{krp@fuw.edu.pl}

\affiliation{Institute of Theoretical Physics, 
             University of Warsaw,
             Ho\.{z}a 69, 00-681 Warsaw, Poland}

\date{\today}

\begin{abstract}
The magnetic shielding $\sigma$ of $^3$He is studied.
The complete relativistic corrections of order $O(\alpha^2)$,
leading QED corrections of order $O(\alpha^3\,\ln\alpha)$, and finite nuclear 
mass effects of order $O(m/m_{\rm N})$ are calculated with 
high numerical precision. The resulting theoretical predictions
for $\sigma = 59.967~43(10)\cdot 10^{-6}$ are the most accurate to date 
among all elements and support the use of $^3$He as a NMR standard.
\end{abstract}

\maketitle

\section{Introduction} 

The coupling of the nuclear magnetic moment $\vec\mu$ 
to an external magnetic field $\vec B$ in closed shell atoms 
is shielded by electrons and thus is slightly smaller
in comparison to the free nucleus \cite{ramsey,chemrev}. 
This shielding is described by the  dimensionless constant $\sigma$
\begin{equation}
H = -\vec \mu\cdot\vec B\,(1-\sigma)\,. \label{01}
\end{equation} 
For a specified atom $\sigma$ is a function of 
the fine structure constant $\alpha$ and depends also 
on the electron-nucleus mass ratio. Since this ratio is
very small, (for $^3$He it is about $ 1.8\cdot 10^{-4}$), 
$\sigma$ can be expanded in a power series in $m/m_{\rm N}$. 
We demonstrate in this work
that the leading term in the mass ratio is numericaly significant,
amounting for $^3$He to $-3.7\cdot 10^{-4}\,\sigma$,
which is 42\% of the relativistic correction.  
Regarding dependence on $\alpha$,
as long as the nuclear charge $Z$ is sufficiently small, say $Z\leq 10$,
the expansion in $\alpha$ is also well convergent. We have therefore for 
$\sigma$ a double series expansion
\begin{equation}
\sigma = \sigma\Bigl(\alpha,\frac{m}{m_{\rm N}}\Bigr) = 
\alpha^2\,\sigma^{(2)} + \alpha^4\,\sigma^{(4)} 
+ \alpha^5\,\sigma^{(5)} +  \alpha^2\,\frac{m}{m_{\rm N}}\sigma^{(2,1)} + \ldots
\label{02}
\end{equation}
The first term of this expansion, $\sigma^{(2)}$, is obtained from
the Ramsey nonrelativistic theory \cite{ramsey} of the magnetic shielding
and for atomic systems takes the very simple form shown in Eq. (\ref{11}).
The derivation of the next coefficient $\sigma^{(4)}$ was considered in a series
of works by Vaara and collaborators \cite{perturb}. They expressed $\sigma^{(4)}$
in terms of the first, second, and third order expectation values
of certain operators with the nonrelativistic wave function.
Numerical evaluations of $\sigma^{(4)}$ were performed for various elements,
but somehow not for $^3$He, for example by Ruud {\em et al.} \cite{ruud}. 
These calculations were not complete, in the sense
that the authors omitted some terms which correspond to $Q_5,Q_8,Q_{11}$ from
our Table \ref{TBL1}, all of them come from the Breit interaction (the second term
in Eq. (\ref{07})). These terms are small, but nevertheless important 
for the estimation of theoretical uncertainties. 
Moreover, the inclusion of the exact electron
g-factor instead of the factor $2$ by the authors of Ref. \cite{perturb}, in our opinion, 
is incorrect, as we explain in Sec. III, devoted to QED effects. 
In general, the relativistic correction $\sigma^{(4)}$ can be obtained from the
Breit-Pauli Hamiltonian in Eq. (\ref{05}), 
in a very similar way to  $\sigma^{(2)}$. In this work,
apart from evaluation of the complete $\sigma^{(4)}$ and $\sigma^{(2,1)}$, 
we present the calculation of the leading logarithmic QED correction  $\sigma^{(5)}$, 
which has a numerical value of about $10^{-5}\,\sigma$.
Finally we consider further improvement of theoretical predictions 
for $^3$He and the other light atomic systems.    

\section{Relativistic correction}
The relativistic correction $\sigma^{(4)}$ can be derived from
the generalized Breit-Pauli Hamiltonian, which 
in addition to relativistic corrections includes coupling to the external
magnetic field \cite{lwqed}. In our case, this field consists of the magnetic field 
$\vec A_I$ coming from the magnetic moment of the nucleus 
\begin{equation}
\vec A_I = \frac{1}{4\,\pi}\,\vec\mu\times\frac{\vec r}{r^3}\,, \label{03}
\end{equation}
and of the homogenous external magnetic field $\vec A_E$
\begin{equation}
\vec A_E = \frac{1}{2}\,\vec B\times\vec r\,. \label{04}
\end{equation}
This Breit-Pauli Hamiltonian in natural units with the external magnetic field $\vec A$
and with $g$ being the electron g-factor is \cite{lwqed}
\begin{eqnarray}
H_{BP} &=& \sum_a H_a + \sum_{a>b,b} H_{ab}\,, \label{05} \\
H_a &=& \frac{\vec\pi_a^2}{2\,m}-\frac{Z\,\alpha}{r_a}
-\frac{\vec\pi^4_a}{8\,m^3}
- \frac{e}{2\,m}\,g\,\vec s_a\cdot\vec B_a  
-\frac{e^2}{8\,m^3}\,\vec B_a^2
+\frac{\pi\,Z\,\alpha}{2\,m^2}\,\delta(\vec r_a)
\nonumber \\ &&
+\frac{e}{8\,m^3}\,\Bigl[2\,\{\vec\pi_a^2 \,,\, \vec s_a\cdot\vec B_a\}
+(g-2)\,\bigl\{\vec\pi_a\cdot\vec B_a\,,\,\vec\pi_a\cdot\vec s_a\bigr\}\Bigr]
\label{06}\\ 
H_{ab} &=& \frac{e^2}{4\,\pi}\,\biggl\{
\frac{1}{r_{ab}} -\frac{1}{2\,m^2}\,\pi_a^i\,
\biggl(\frac{\delta^{ij}}{r_{ab}}+\frac{r^i_{ab}\,r^j_{ab}}{r^3_{ab}}
\biggr)\,\pi_b^j 
+\frac{\pi}{m^2}\,\delta(\vec r_{ab})
\nonumber \\ &&
+\frac{g^2}{4\,m^2}\,\frac{s_a^i\,s_b^j}{r_{ab}^3}\,
\biggl(\delta^{ij}-3\,\frac{r_{ab}^i\,r_{ab}^j}{r_{ab}^2}\biggr)
+\frac{1}{2\,m^2\,r_{ab}^3} \biggl[
g\,\vec s_a\cdot\vec r_{ab}\times\vec\pi_b 
\nonumber \\ &&
-g\,\vec s_b\cdot\vec r_{ab}\times\vec\pi_a +
(g-1)\,\vec s_b\cdot\vec r_{ab}\times\vec\pi_b 
-(g-1)\,\vec s_a\cdot\vec r_{ab}\times\vec\pi_a\bigr)\biggr]\biggr\}\,.
\label{07}
\end{eqnarray}
where $\vec \pi = \vec p-e\,\vec A$, and $H_{BP}$ in the above includes 
dependence on the electron g-factor only for spin dependent terms.
For the derivation of $\sigma^{(4)}$
we set $g=2$ and separate  $H_{BP}$ into parts: the leading interaction
with the external field, no external field $H_{BP}\bigr|_{\vec A_E=\vec A_I=0}$,
linear in the homogenous field $\delta_{\vec A_E}H_{BP}\bigr|_{\vec A_I=0}$,
linear in the nuclear magnetic field $\delta_{\vec A_I}H_{BP}\bigr|_{\vec A_E=0}$,
and bilinear in the homogenous and the nuclear magnetic fields 
$\delta_{\vec A_E,\vec A_I}H_{BP}$.
 \begin{eqnarray}
H_{BP} &=& \frac{e^2}{m}\,\vec A_E\cdot\vec A_I 
-\frac{e}{2\,m}\,\sum_a(\vec L_a+2\,\vec s_a)\cdot\vec B
\nonumber \\ &&
+ H_{BP}\bigr|_{\vec A_E=\vec A_I=0} 
+ \delta_{\vec A_E}H_{BP}\bigr|_{\vec A_I=0}
+ \delta_{\vec A_I}H_{BP}\bigr|_{\vec A_E=0}
+ \delta_{\vec A_E,\vec A_I}H_{BP} \label{08}
\end{eqnarray}
Corrections to the energy, which are bilinear in magnetic fields, 
can be represented in terms of the leading contribution $E^{(2)}$ and 
the relativistic correction $E^{(4)}$,
\begin{eqnarray}
E^{(2)} &=&  \frac{e^2}{m}\,\langle\vec A_E\cdot\vec A_I\rangle \label{09} \\
E^{(4)} &=& 2\,\biggl\langle\frac{e^2}{m}\,\vec A_E\cdot\vec A_I
\frac{1}{(E-H)'}\,H_{BP}\bigr|_{\vec A_E=\vec A_I=0}\biggr\rangle \nonumber\\ &&
+2\,\biggl\langle \delta_{\vec A_E}H_{BP}\bigr|_{\vec A_I=0}\,
\frac{1}{(E-H)'}\,\delta_{\vec A_I}H_{BP}\bigr|_{\vec A_E=0}\biggr\rangle\nonumber \\ &&
+ \bigl\langle \delta_{\vec A_E,\vec A_I}H_{BP} \bigr\rangle\,. \label{10}
\end{eqnarray}
Because of the spherical symmetry of closed shell atoms,
perturbations due to $-\frac{e}{2\,m}\,\sum_a(\vec L_a+2\,\vec s_a)\cdot\vec B$ vanish,
which is a significant simplification over molecular systems.
Thus, in the leading order the shielding constant $\sigma$ takes the form
\begin{equation}
\sigma^{(2)} = \frac{1}{3}\,\sum_a\biggl\langle\frac{1}{r_a}\biggr\rangle, \label{11}
\end{equation}
while relativistic corrections are
\begin{eqnarray}
\sigma^{(4)} &=& \sigma_1^{(4)} + \sigma_{2A}^{(4)} + \sigma_{2B}^{(4)} +
\sigma_{2C}^{(4)} + \sigma_3^{(4)} \label{12}\\
\sigma_1^{(4)} &=& \frac{2}{3}\,\biggl\langle\biggl(\frac{1}{r_1}+\frac{1}{r_2}\biggr)\,
\frac{1}{(E-H)'}\,\biggl[\sum_a \biggl(\frac{\pi\,Z}{2}\,\delta(\vec r_a)-\frac{p_a^4}{8}\biggr)
+\pi\,\delta(\vec r)-\frac{1}{2}\,p_1^i\biggl(\frac{\delta^{ij}}{r}+\frac{r^i\,r^j}{r^2}\biggr)\,p_2^j
\biggr]\biggr\rangle \nonumber \\ \label{13}\\
\sigma_{2A}^{(4)} &=&-\frac{2}{9}\,\biggl\langle\pi\,\bigl[\delta(\vec
  r_1)-\delta(\vec r_2)\bigr]\,
\frac{1}{(E-H)}\,\biggl[
3\,p_1^2-3\,p_2^2-\frac{Z}{r_1}+\frac{Z}{r_2}-\frac{\vec r\cdot(\vec r_1+\vec r_2)}{r^3}
\biggr]\biggr\rangle \label{14}\\
\sigma_{2B}^{(4)} &=& -\frac{1}{6}\,\biggl\langle\biggl(
\frac{\vec r_1\times\vec p_1}{r_1^3} + \frac{\vec r_2\times\vec p_2}{r_2^3}\biggr)\,
\frac{1}{(E-H)}\nonumber \\ && \biggl[
\vec r_1\times\vec p_1\,p_1^2 + \vec r_2\times\vec p_2\,p_2^2  
+\frac{1}{r}\,\vec r_1\times\vec p_2 + \frac{1}{r}\,\vec r_2\times\vec p_1
-\vec r_1\times\vec r_2\,\frac{\vec r}{r^3}\cdot(\vec p_1+\vec p_2)\biggr]\biggr\rangle \label{15}\\
\sigma_{2C}^{(4)} &=& -\frac{1}{8}\,\biggl\langle
\biggl(\frac{r_1^i\,r_1^j}{r_1^5}-\frac{r_2^i\,r_2^j}{r_2^5}\biggr)^{(2)}\,
\frac{1}{(E-H)}\,\biggl(Z\,\frac{r_1^i\,r_1^j}{r_1^3} - Z\,\frac{r_2^i\,r_2^j}{r_2^3}
+\frac{r^i}{r^3}\,(r_1^j+r_2^j)\biggr)^{(2)}\biggr\rangle \label{16}\\
\sigma_3^{(4)} &=& 
\frac{1}{12}\,\biggl\langle
\biggl(\frac{1}{r_1^3}+\frac{1}{r_2^3}\biggr)\,
\biggl(\frac{\vec r\cdot\vec r_1\;\vec r\cdot\vec r_2}{r^3} - 3\,\frac{\vec r_1\cdot\vec r_2}{r}\biggr)
\biggr\rangle
\nonumber \\ &&
-\frac{1}{6}\,\sum_a\biggl\langle
 \frac{1}{r_a}\,p_a^2+\frac{(\vec r_a\times\vec p_a)^2}{r_a^3}+4\,\pi\,\delta(\vec r_a)
\biggr\rangle \label{17}
\end{eqnarray}
where $\vec r\equiv \vec r_1-\vec r_2$, $(p^i\,q^i)^{(2)} =
p^i\,q^j/2+p^j\,q^i/2-\delta^{ij}\,\vec p\cdot\vec q/3$,
and $1/(E-H)'$ is the reduced Green function (the reference state is subtracted out).
We have split these relativistic corrections into first order terms
$\sigma_3^{(4)}$, the second order terms with intermediate singlet $^1S$- 
$\sigma_1^{(4)}$, triplet $^3S$- $\sigma_{2A}^{(4)}$, singlet $^1P^e$- $\sigma_{2B}^{(4)}$, 
and triplet $^3D$- $\sigma_{2C}^{(4)}$. These terms form a complete relativistic
correction of order $O(\alpha^4)$ and  their numerical calculations are described in Sec. V.

\section{QED effects}
The next order correction $O(\alpha^5)$ comes from QED effects.
They contribute by $F_i$ electromagnetic formfactors \cite{itzykson}, 
the magnetic susceptibility and the so called Bethe logarithms.  
The slope of $F'_1(0)$ and $F_2(0)=(g-2)/2$
are known analytically at one-loop order \cite{itzykson}. However,  
$F'_1(0)$ is infrared divergent and this divergence cancels out with the uv divergence from 
the low-energy contribution in a similar way as for the Lamb shift in hydrogen \cite{itzykson}.
The contribution from $F_2(0)$ is encoded 
by the g-factor in the Breit-Pauli Hamiltonian $H_{BP}$, Eqs. (\ref{05}-\ref{07}). 
We note that the g-factor enters relativistic corrections with different
coefficients, which is not in accordance with Ref. \cite{ruud}.
The radiative corrections to the magnetic susceptibility have not yet been
evaluated, but their calculation can be performed along the lines of Ref. \cite{glong}.
Finally the Bethe logarithmic contribution, as for the Lamb shift,
is probably the most difficult part of the numerical evaluation and
can be obtained probably only for simple systems. However, 
the total QED correction can easily be estimated on the basis of the leading
logarithmic contribution, which is derived below.     

The leading logarithmic correction
to $\sigma$ can be obtained in the same way as with the Lamb shift.
One considers 2-electron self interaction in the magnetic field
due to low-energy photons  ($m=1$)
\begin{equation}
E_L = e^2\,\int^\epsilon\frac{d^3k}{(2\,\pi)^3\,2\,k}\,
\biggl(\delta^{ij}-\frac{k^i\,k^j}{k^2}\biggr)\,
\biggl\langle\phi\biggl|(\pi_1^i+\pi_2^i)\,
\frac{1}{E-H-k}\,(\pi_1^j+\pi_2^j)\biggr|\phi\biggr\rangle 
\label{18}
\end{equation}
where $\phi$ is the eigenstate of the nonrelativistic Hamiltonian $H$ 
in the magnetic field, with energy $E$.  The logarithmic part is
\begin{eqnarray}
E_{\rm Log} &=& \frac{\alpha}{3\,\pi}\,\ln\bigl[(Z\,\alpha)^{-2}\bigr]\,
\langle\phi|[\vec\pi_1+\vec \pi_2\,,\,
[H-E\,,\,\vec \pi_1+\vec \pi_2]]|\phi\rangle 
\nonumber \\ &=&
\frac{\alpha}{3\,\pi}\,\ln\bigl[(Z\,\alpha)^{-2}\bigr]\,
\sum_a \biggl\langle\phi\biggl|
4\,\pi\,Z\,\alpha\,\delta(\vec r_a)+2\,e^2\,B_a^2+\frac{e}{2}\,
\bigl\{\vec \pi_a\,,\,\nabla_a\times\vec B_a\bigr\}
\biggr|\phi\biggr\rangle \label{19}
\end{eqnarray}
The resulting logarithmic contribution to the
shielding constant in order $\alpha^5$ is 
\begin{eqnarray}
\sigma^{(5)} &=&\frac{8\,Z}{9}\,\ln\bigl[(Z\,\alpha)^{-2}\bigr]\,
\biggl\langle\biggl(\frac{1}{r_1}+\frac{1}{r_2}\biggr)\,\frac{1}{(E-H)'}\,
\bigl[\delta(\vec r_1)+\delta(\vec r_2)\bigr]\biggr\rangle
\nonumber \\ &&
+\frac{20}{9}\,\ln\bigl[(Z\,\alpha)^{-2}\bigr]\,
\bigl\langle\delta(\vec r_1)+\delta(\vec r_2)\bigr\rangle
\nonumber\\ &&
+\frac{28}{9}\,\ln\alpha\,\biggl\langle\delta(\vec r)\,\frac{1}{(E-H)'}\,
\biggl(\frac{1}{r_1}+\frac{1}{r_2}\biggr)\biggr\rangle \label{20}
\end{eqnarray}
where we have added a small second order $\ln\alpha$ term,
which comes from the two-electron Lamb shift \cite{sap}, and thus Eq. (\ref{20}) 
is a complete logarithmic contribution.

\section{Nuclear mass corrections}
Effects coming from the finite nuclear mass are usually, if not always,
neglected. Here we use the result derived in our previous work
\cite{magnetic}, namely the leading correction $\sigma^{(2,1)}$ is given by 
\begin{eqnarray}
\sigma^{(2,1)} &=& \frac{1}{3}\,\biggl\langle\sum_a\frac{1}{r_{a}}\,
\frac{1}{(E-H)'}\,p_N^2\biggr\rangle
+\frac{1}{3}\,\frac{(1-g_N)}{Z\,g_N}\,\langle p_N^2\rangle \label{21}\\&&
+\frac{1}{3}\,\biggl\langle
\bigl(\vec r_1\times\vec p_2 +\vec r_2\times\vec p_1\bigr)\,
\frac{1}{(E-H)}\,
\sum_{a}\,\frac{\vec r_{a}}{r_a^3}\times \vec p_a\biggr\rangle
\nonumber
 \end{eqnarray}
where $\vec p_N = -\sum_a \vec p_a$, and
\begin{equation}
g_{\rm N} = \frac{m_{\rm N}}{Z\,m_{\rm p}}\,\frac{\mu}{\mu_{\rm n}}\,\frac{1}{I}, \label{22}
\end{equation}
where $\mu_{\rm n}$ is the nuclear magneton, $I$ is the nuclear spin, and
$m_p$ is the proton mass. The definition of the nuclear g-factor assumed here
is different from the standard one by the use of 
the actual charge $e_N$ and the mass $m_N$ of the corresponding
particle, namely the coupling of the spin to the magnetic 
field is $-e_N\,g_N/(2\,m_N)\,\vec I\cdot\vec B$.
The numerical value of $g_N$ obtained
from the known spin $I=1/2$, nuclear charge $Z=2$, magnetic moment 
$\mu =-2.127~625~2(1)\,\mu_{\rm n}$, and the mass ratio 
$m_{\rm N}/m_{\rm p} = 2.993~152~671~3(26)$
is presented in the caption of Table \ref{TBL2}, while the numerical evaluation
of $\sigma^{(2,1)}$ is performed in Sec. V. Our result is not in agreement
with the work on Neronov and Barzakh in \cite{rus}, 
which based on earlier results of Hegstrom in \cite{hegstrom}. 

\section{Numerical calculations}
The numerical calculations are performed with the use of
explicitly correlated exponential functions,
which for the S-state have the form
\begin{equation}
\phi(r_1,r_2,r) = \sum_{i=1}^{\cal N} v_i
[e^{-\alpha_i r_1-\beta_i r_2-\gamma_i r} \pm (r_1 \leftrightarrow r_2)],
\label{23}
\end{equation}
where $\alpha_i$, $\beta_i$ and $\gamma_i$ are generated randomly
with conditions:
\begin{eqnarray}
A_1<\alpha_i<A_2,\;\; \beta_i+\gamma_i>\varepsilon, \nonumber \\
B_1<\beta_i<B_2,\;\; \alpha_i+\gamma_i>\varepsilon, \label{24}\\
C_1<\gamma_i<C_2,\;\; \alpha_i+\beta_i>\varepsilon. \nonumber
\end{eqnarray}
with $\varepsilon$ approximately equal to $\sqrt{2\,m\,(E_{\rm He^+}-E_{\rm He})}$. 
In order to obtain a more accurate wave function, following
Korobov \cite{kor}, we use double set of the form (\ref{24}). 
Parameters $A_i,B_i,C_i$ are determined 
by minimization of the nonrelativistic energy.
The linear coefficients $v_i$ in Eq. (\ref{23}) are obtained from
a solution of the generalized eigenvalue problem with the length 
of the basis set ${\cal N} = 100, 300, 600, 900, 1200, 1500$ 
using extended precision arithmetic. 
As a result we obtain the following nonrelativistic energy in au
\begin{equation}
E_0(1^1S_0) = -2.903~724~377~034~119~593(5), \label{25}
\end{equation}
in agreement with the even more accurate result of 
Korobov \cite{kor} and of Drake in \cite{Drake_h}.
The calculation of matrix elements of the nonrelativistic Hamiltonian
are performed with the use of a simple formula for the master integral:
\begin{equation}
\frac{1}{16\,\pi^2}\,\int d^3 r_1\,\int d^3r_2\,
\frac{e^{-\alpha r_1-\beta r_2-\gamma r}}{r_1\,r_2\,r} =
\frac{1}{(\alpha+\beta)(\beta+\gamma)(\gamma+\alpha)}. \label{26}
\end{equation}
Integrals with any additional powers of $r_i$ in the numerator can be 
obtained by differentiation with respect to the corresponding parameter 
$\alpha$, $\beta$ or $\gamma$. Matrix elements of relativistic corrections
involve inverse powers of $r_1, r_2, r$, and these can be obtained 
by integration with respect to the corresponding parameter.
In fact, all matrix elements involved in relativistic, QED and the finite nuclear mass
corrections  can be expressed in terms of rational, logarithmic and
dilogarithmic functions in $\alpha,\beta$, and $\gamma$.
Considering numerical convergence, first order matrix elements can be
calculated  accurately using even short, i.e. ${\cal N}=300$ expansion of 
the nonrelativistic wave function.
The calculation of the second order corrections $Q_{6-13,15-16}$ is more complicated.
The inversion of the operator $E-H$ 
is performed in a basis set of even parity functions with $l=0,1,2$ of the form
in Eq. (\ref{23}) and  
\begin{eqnarray}
\vec \phi(r_1,r_2,r) &=& \sum_k v_k\,\vec r_1 \times \vec r_2\,
[e^{-\alpha_k r_1-\beta_k r_2-\gamma_k r} - (r_1 \leftrightarrow r_2)]\,,\label{27}\\
\phi^{ij}(r_1,r_2,r) &=& \sum_k v_k\, [(r_1^i\,r_1^j-r_1^2\,\delta^{ij}/3)\,
e^{-\alpha_k r_1-\beta_k r_2-\gamma_k r}- (r_1 \leftrightarrow r_2)] \nonumber
\\ && +\sum_l v_l\, (r^i\,r^j-r^2\,\delta^{ij}/3)\,
[e^{-\alpha_l r_1-\beta_l r_2-\gamma_l r}- (r_1 \leftrightarrow r_2)] \label{28}
\end{eqnarray}
The values of parameters $A_i$, $B_i$ and $C_i$
are obtained by minimization of the appropriate functional,
which is the symmetric second order matrix element with the operator
standing on the right hand side of the corresponding $Q_i$ in Table I. 
This explicitly  correlated exponential basis set 
allows us to obtain precise matrix elements
of all $Q_i$ operators, and the numerical results are presented in Table I.
Some of these matrix elements have already been presented in \cite{Drake_h},
and results in Table I are in agreement with them.

\begin{table}[htb]
\caption{ Expectation values of operators entering $\sigma$, 
          all digits are significant, $\vec r= \vec r_1 - \vec r_2 $, and
          $1/(E-H)$ is the nonrelativistic Green function.}
\label{TBL1}
%\begin{tabular}{l@{\hspace{0.2cm}}.@{\hspace{1.0cm}}.}
%\begin{tabular}{ll@{\hspace{0.1cm}}.}
\begin{tabular}{ll@{\hspace{-1.0cm}}.}
\hline
\hline
%\rule[-3mm]{0mm}{8mm}
%&
%\multicolumn{1}{l}{operator} &
%\multicolumn{1}{c}{expectation value}  \\
%\hline
$Q_{1}=$ &$\frac{1}{r_1}+\frac{1}{r_2}        $
         & 3x.376~633~601 \\
$Q_{2}=$ &$\frac{1}{r_1}\,p_1^2+\frac{1}{r_2}\,p_2^2 $
         &  33x.677~743 \\
$Q_{3}=$ &$\frac{(\vec r_1\times\vec p_1)^2}{r_1^3}+\frac{(\vec r_2\times\vec p_2)^2}{r_2^3} $
         & 0x.073~109 \\
$Q_{4}=$ &$\delta(\vec r_1)+\delta(\vec r_2) $
         & 3x.620~859 \\
$Q_{5}=$ &$\bigl(\frac{1}{r_1^3}+\frac{1}{r_2^3}\bigr)\,
          \bigl(\frac{\vec r\cdot\vec r_1\;\vec r\cdot\vec r_2}
          {r^3}-3\frac{\vec r_1\cdot\vec r_2}{r}\bigr)$
         &  -3x.435~251 \\
$Q_{6}=$ &$[\delta(\vec r_1)+\delta(\vec r_2)]
          \,\frac{1}{(E-H)'}\,
          \bigl(\frac{1}{r_1}+\frac{1}{r_2}\bigr) $
         & -3x.025~857 \\
$Q_{7}=$ &$(p_1^4+p_2^4)\,\frac{1}{(E-H)'}\,
          \bigl(\frac{1}{r_1}+\frac{1}{r_2}\bigr) $
         &  -118x.140~232 \\
$Q_{8}=$ &$p_1^i\,\bigl(\frac{\delta^{ij}}{r}+\frac{r^i\,r^j}{r^3}\bigr)\,p_2^j
           \,\frac{1}{(E-H)'}\,\bigl(\frac{1}{r_1}+\frac{1}{r_2}\bigr) $
         &  -0x.208~449 \\
$Q_{9}=$ &$\delta(\vec r)\,\frac{1}{(E-H)'}\,
          \bigl(\frac{1}{r_1}+\frac{1}{r_2}\bigr) $
         &  -0x.112~964 \\
$Q_{10}=$&$(p_1^2\,\vec r_1\times\vec p_1+p_2^2\,\vec r_2\times\vec p_2)\,\frac{1}{E-H}\,
         \bigl(\frac{\vec r_1\times\vec p_1}{r_1^3}+\frac{\vec r_2\times\vec p_2}{r_2^3}\bigr)$
         & -0x.041~690 \\
$Q_{11}=$&$\Bigl[\frac{1}{r}\,(\vec r_2\times\vec p_1+\vec r_1\times\vec p_2)
         -\vec r_1\times\vec r_2\,\frac{\vec r}{r^3}\cdot(\vec p_1+\vec p_2)\Bigr]
         \,\frac{1}{E-H}\,\bigl(\frac{\vec r_1\times\vec p_1}{r_1^3}
         +\frac{\vec r_2\times\vec p_2}{r_2^3}\bigr)$
         &  0x.037~546 \\
$Q_{12}=$&$\bigl(\frac{r_1^i\,r_1^j}{r_1^5}-\frac{r_2^i\,r_2^j}{r_2^5}\bigr)^{(2)}
          \,\frac{1}{E-H}\,\bigl(Z\,\frac{r_1^i\,r_1^j}{r_1^3}-Z\,\frac{r_2^ir_2^j}{r_2^3}
          +\frac{r^i\,(r_1^j+r_2^j)}{r^3}\bigr)^{(2)}$
          &  -6x.838~1(4) \\
$Q_{13}=$&$[\delta(\vec r_1)-\delta(\vec r_2)]\,\frac{1}{E-H}\,
          \bigl(3\,p_1^2-3\,p_2^2-\frac{Z}{r_1}+\frac{Z}{r_2}-\frac{r_1^2-r_2^2}{r^3}\bigr)$
          &  -39x.921~269 \\
$Q_{14}=$ &$(\vec p_1+\vec p_2)^2 $
          &  6x.125~588 \\
$Q_{15}=$&$(\vec p_1+\vec p_2)^2\,\frac{1}{(E-H)'}\,\bigl(\frac{1}{r_1}+\frac{1}{r_2}\bigr)$
         &  -3x.504~997 \\
$Q_{16}=$&$(\vec r_1\times\vec p_2+\vec r_2\times\vec p_1)\,
           \frac{1}{E-H}\,
           \bigl(\frac{\vec r_1\times\vec p_1}{r_1^3}
                +\frac{\vec r_2\times\vec p_2}{r_2^3}\bigr) $
         &  0x.078~743 \\[1ex]
\hline
\hline
\end{tabular}
%\end{minipage}
\end{table}

\section{Results}
All corrections to the shielding constant $\sigma$ can be expressed in terms
of $Q_i$ values and results are presented in Table II.
\begin{table}[htb]
\caption{ Contribution to the shielding constant. Physical
  constants are taken from \cite{nist}: $\alpha^{-1} = 137.035~999~679(94),
  g_{\rm N} = -6.368~307~2,
  m_{\rm N}/m =5~495.885~276~5(52)$. Uncertainty of $\sigma$ is set to 
  20\% of $\sigma^{(5)}$ contribution}
\label{TBL2}
%\begin{tabular}{l@{\hspace{0.2cm}}.@{\hspace{1.0cm}}.}
%\begin{tabular}{ll@{\hspace{0.1cm}}.}
%\begin{tabular}{rl@{\hspace{-2.0cm}}.}
\begin{tabular}{rl@{\hspace{-1.0cm}}.@{\hspace{-1.0cm}}.}
\hline
\hline
%\rule[-3mm]{0mm}{8mm}
&
\multicolumn{1}{l}{operator} &
\multicolumn{1}{c}{expectation value} &
\multicolumn{1}{c}{contribution to $\sigma\times 10^6$}  \\
\hline
$\sigma^{(2)} =$&$ \frac{1}{3}\,Q_1$ & 1x.125~544~534 & 59x.936~770  \\
$\sigma_1^{(4)} =$&$\frac{\pi Z}{3}\,Q_6 -\frac{1}{12}\,Q_7
-\frac{1}{3}\,Q_8 +\frac{2\pi}{3}\,Q_9$ & 3x.340~6 & \\
$\sigma_{2A}^{(4)} =$&$-\frac{2\pi}{9}\,Q_{13}$& 27x.870~3 &  \\
$\sigma_{2B}^{(4)} =$&$-\frac{1}{6}\,Q_{10} -\frac{1}{6}\,Q_{11}$&0x.000~7 & \\
$\sigma_{2C}^{(4)} =$&$-\frac{1}{8}\,Q_{12}$&0x.854~8 & \\
$\sigma_3^{(4)} =$&$ -\frac{1}{6}\,Q_2-\frac{1}{6}\,Q_3-\frac{2\pi}{3}\,Q_4
                  +\frac{1}{12}\,Q_5 $&-13x.494~9 & \\
$\sigma^{(4)} =$&$\sigma_1^{(4)} + \sigma_{2A}^{(4)} + \sigma_{2B}^{(4)} +
                  \sigma_{2C}^{(4)} + \sigma_3^{(4)}$
                &18x.571~4 & 0x.052~663\\
$\sigma^{(5)}=$&$\ln\bigl[(Z\,\alpha)^{-2}\bigr]\,\Bigl(\frac{8\,Z}{9}\,Q_6 
              + \frac{20}{9}\,Q_4\Bigr)
              + \frac{28}{9}\,\ln\alpha\;Q_9$&24x.277~0 & 0x.000~502\\
$\sigma^{(2,1)} =$&$
\frac{1}{3}\,\Bigl(\frac{1-g_\mathrm{N}}{Zg_\mathrm{N}}\,Q_{14}+Q_{15}+Q_{16}\Bigr)$ 
                &-2x.323~3 & -0x.022~511\\
$\sigma =$&$\alpha^2\,\sigma^{(2)} + \alpha^4\,\sigma^{(4)} 
+ \alpha^5\,\sigma^{(5)} +  \alpha^2\,\frac{m}{m_{\rm N}}\,\sigma^{(2,1)}$&&59x.967~43(10)
\\[1ex]
\hline
\hline
\end{tabular}
%\end{minipage}
\end{table}
All of them are accurate to all digits shown, nevertheless the uncertainty is
different from 0, due to the neglect of the non logarithmic part of $\sigma^{(5)}$
which we estimate, on the basis of the helium Lamb shift, to be about 20\%.
The relativistic correction $\sigma^{(4)}$ is relatively large, namely $10^{-3}$
of the nonrelativistic one, and is dominated by the second order contribution
from the triplet $S$ states. QED corrections are non negligible, 1\% of 
the relativistic contribution, while the finite nuclear mass corrections are very 
significant, about 42\% of the relativistic contribution and of the opposite
sign. Except for neglected QED contributions, in our opinion,
no other correction including the finite nuclear size,
may alter the result at the 0.1 ppb level.

\section{Summary}
We have obtained relativistic, QED and finite nuclear mass
corrections to the magnetic shielding constant in $^3$He
with the uncertainty of 0.1 ppb, which is caused by neglected QED corrections.
While our nonrelativistic result is in perfect agreement
with the previous one by Drake in \cite{Drake_h}, the relativistic
correction $0.052\,663$ is in the moderate agreement with result 
by Vaara and Pyykk\"o in Ref. \cite{vaara} $0.04$, which is obtained
as the difference between DF LR and HF values in their Table I.
We think therefore, that our calculation needs more accurate confirmation,
since we have not been able to test individual relativistic corrections.
However, under the assumption that the present calculation is correct, the shielding
factor for $^3$He is now known with the highest accuracy of any atom, which supports
its use as a NMR standard. Moreover, the theoretical accuracy can be further
improved by the complete calculation of the QED effects, which for He
is certainly possible. Regarding calculations for other light atoms
and molecules, we are not convinced that the commonly used Gaussian functions
can be applied for the accurate evaluation of second order matrix elements,
especially for $Q_7,\,Q_{12}$ and $Q_{13}$. It is likely that the use of linear terms,
which improve the cusp condition of the nonrelativistic wave function,
will be necessary in order to obtain numerical result with predictable
uncertainty. If this can be achieved, it will open a window
for high accuracy determination of nuclear magnetic moments.
Since the NMR frequencies can be measured very accurately,
as that in $^3$He, with respect to the proton in tetramethylsilane
(TMS) \cite{3he}, the magnetic moment of helion can be related
to the accurately measured proton magnetic moment \cite{nist}. 

\section*{Acknowledgments}
This work was supported by NIST
through Precision Measurement Grant PMG 60NANB7D6153.

\end{document}